\documentclass[12pt]{iopart}

\usepackage{color}  
\expandafter\let\csname equation*\endcsname\relax
\expandafter\let\csname endequation*\endcsname\relax
\usepackage{amsmath}
\usepackage{graphicx}
\usepackage{soul}
\usepackage[export]{adjustbox}
\usepackage{physics}
\usepackage{amssymb}
\usepackage{xcolor}
\usepackage{url}
\usepackage[hyperindex,breaklinks]{hyperref}

\begin{document}

\title[Classical Geometric Fluctuation Relations]{Classical Geometric Fluctuation Relations}

\author{Pedro B. Melo$^{1}$, S\'ilvio M. Duarte Queir\'os$^{2,3}$, Diogo O. Soares-Pinto$^{4}$, Welles A. M. Morgado$^{1,3}$}

\address{$^{1}$Departamento de Física, Pontifícia Universidade Católica, 22452-970, Rio de Janeiro RJ, Brazil}
\address{$^{2}$ Centro Brasileiro de Pesquisas Físicas, 22290-180, Rio de Janeiro RJ, Brazil}
\address{$^{3}$ INCT-Sistemas Complexos, Brazil} 
\address{$^{4}$ Instituto de Física de São Carlos, Universidade de São Paulo, 13560-970 São Carlos SP, Brazil}
\ead{pedrobmelo@aluno.puc-rio.br}
\vspace{10pt}
\begin{indented}
\item[]March 2025
\end{indented}


\begin{abstract}
Fisher Information (FI) is a quantity ubiquitously measured in such varied areas like metrology, machine learning, and biological complexity. Mathematically, it represents a lower bound in the variance of unknown parameters that are related to the distributions one has access, and a metric for probability manifolds. A stochastic analogous of the Fisher Information, dubbed stochastic Fisher Information was recently introduced in the literature by some of us. By exploring the probability distributions of the Stochastic Fisher Information (SFI), we uncover two fluctuation relations with an inherent geometric nature, as the SFI acts as a single nonequilibrium trajectory metric. The geometric nature of these relations is expressed through a stochastic length in entropy space derived from the system entropy associated with a nonequilibrium trajectory. We also explore the possibility of trajectory-dependent uncertainty relations linked to the SFI with time as a parameter.
Finally, we test our geometric fluctuation relations using two nonequilibrium models.
\end{abstract}

%
\vspace{2pc}
\submitto{\JPA}
%
%
%

\section{Introduction} Among the successes of the stochastic thermodynamic theory are the Fluctuation theorems (FT) \cite{Evans1993, Gallavoti1995, jarzynski1997, Crooks1999}. For instance, by recasting the second law of thermodynamics with statistical formulation, one connects the average work made out of equilibrium to the variation of equilibrium free energy. This theorem is known as the Jarzynski integral FT \cite{jarzynski1997}. A stricter, alternative formulation to an FT is the detailed FT \cite{Crooks1999}, where for a trajectory in which involution is valid, the probability distribution of the reverse trajectory is proportional to the direct trajectory probability distribution up to an entropic cost, of the total entropy produced in the direct trajectory~\footnote{By direct and reverse we mean the trajectory that follows the thermodynamic arrow of time.}.

Non-equilibrium thermodynamics can be studied within a geometric approach as well \cite{Weinhold1975, Ruppeiner1979, Salamon1983, Ruppeiner1995, Brody1995, crooks2007} using the definition of distance in probability space. In such framework, quantities such as the Fisher information (FI) \cite{crooks2007} become a metric on probability space. Considering two unknown parameters, $\theta_{i}$ and $\theta_{j}$, given that the probability distribution $P(x|\theta_{i,j})$ is a conditional probability of these parameters, the FI is defined as
\begin{equation}
    \mathcal{I}(\theta_i,\theta_j) \equiv \left\langle\left(\frac{\partial_{\theta_i} P(x|\theta_{i})}{P(x|\theta_{i})}\right)\left(\frac{\partial_{\theta_j} P(x|\theta_{j})}{P(x|\theta_{j})}\right)\right\rangle_{P(x)}. \label{FI_average}
\end{equation}
The FI is a metrological quantity that determines the minimum variance of the parameters $\theta$. At the same time, it represents a metric in probability space that is contractive under stochastic maps. A line element of displacement in the probability space follows
\begin{equation}
    {\rm d}s^2 = \mathcal{I}(\theta_{i},\theta_{j}){\rm d}\theta_{i}{\rm d}\theta_{j}.
\end{equation}
For time-dependent parameters $\theta_{i,j}(t)$, we have 
\begin{equation}
    {\rm d}s^2 = \mathcal{I}(\theta_{i},\theta_{j})\dv{\theta_{j}}{t}\dv{\theta_{i}}{t}~{\rm d}t^{2}.
\end{equation}
The length $\mathcal{L}$ of a path in the probability space is given, using the FI metric, by
\begin{equation}
    \mathcal{L}(\tau) = \int {\rm d}s= \int_{0}^{\tau} {\rm d}t \sqrt{\dv{\theta_{i}}{t}\mathcal{I}(\theta_{i},\theta_{j})\dv{\theta_j}{t}} = \int_{0}^{\tau} {\rm d}t \sqrt{\mathcal{I}(t)}.\label{eq:thermodynamic_length}
\end{equation}

Although the FI is a key concept in many areas such as biology \cite{frank2009}, ecology \cite{mayer2006}, and economy \cite{sahalia2008}, its interest has sparkled within metrological and informational contexts, specially in quantum metrology and information \cite{Escher2011, Toth2012, hyllus2012, Liu_2020, min2022, marvian2022}, where the FI is used as an estimator of an unknown parameter related to a known distribution due to the Cram\'er-Rao (CR) bound \cite{Cramer_1946, Rao1992}. The inverse of FI emerges in this scenario as the lower bound to the variance of the unknown parameters meaning the variance of $\theta$ is lower bound by the inverse of the FI of $\theta$. 

In the context of single-parameter Fisher Information (FI), $\theta_{i} = \theta_{j} = \theta  $, and when $\theta $ is time-dependent, -- \textit{i.e.}, the parameter associated with a time-varying protocol -- Eq.~(\ref{FI_average}) can be reformulated and reads,
\begin{equation}
\mathcal{I}(t) = \iiint {\rm d}x_{0}{\rm d}x_{\tau}\mathcal{D}x~\Dot{\theta}^{2}(t) \left(\pdv{\log P(x|\theta(t))}{\theta(t)}\right)^2 ,
\label{FI_time}
\end{equation}
where the integrals indicate an average on $P(x|\theta(t))$; particularly, when the parameter is the time itself, $\theta(t) = t$, the FI is given by 
\begin{equation}
    \mathcal{I}(t) = \left\langle \left(\pdv{\log P(x,t)}{t}\right)^2 \right\rangle,
\end{equation}
which is the case we consider along this manuscript. In stochastic systems, the entropy of the system is given by the surprisal $s_{{\rm sys}}(x,t) = -\log P(x,t)$ (we consider $k_{B} = 1$ throughout the whole manuscript). It means that this equation shows that FI represents the average stochastic system entropy production. Using the CR bound, one can establish a series of thermodynamic uncertainty relations (TUR) \cite{Nicholson2018, Ito2018, TanvanVU2019_TUR, ItoDechant2020, nicholson2020, Falasco2020} and relates to the concept of speed limits for physical processes as well~\cite{Busch2002}. In classical systems, it has been shown that in steady-state conditions (whether equilibrium or not), there are trade-off between the speed at which a system evolves from one state to another and the entropy produced during that change~\cite{Okuyama2018, Shiraishi2018, Ito2018, Gupta2020}. 

The CR bound is recognized as a generalized trade-off relation \cite{ItoDechant2020}, which recovers known relations for time. Furthermore, minimal entropy production is closely linked to speed limit bounds and TURs \cite{TanVanVu2020}, with these relations holding even for systems far from equilibrium \cite{Salazar2022, TanVanVu2023}. Even though the matter of speed limits is outside of the scope of this paper, our results indicate the existence of such via Cauchy-Schwarz inequality to be discussed elsewhere \cite{melo2025_speedlimits}.

Recently, it has been introduced by some of us a stochastic interpretation of the Fisher Information, coined Stochastic Fisher Information (SFI) \cite{melo2024stochasticthermodynamicsfisherinformation}. This quantity is a random variable that follows a distribution with average equivalent to the FI. The physical interpretation of the SFI represents a metric for fluctuating paths in the probability space, analogous to proposals for gravity \cite{Moffat1997} and an informational stochastic quantity, much like the stochastic entropy \cite{seifert2005}.

In this manuscript, we explore the SFI seeking fluctuation relations (FR) related to such quantity. In Section \ref{sec:2}, we introduce the SFI when time is the parameter and discuss its physical meaning. In Section \ref{sec:3}, we present our first main result, which states that there is a detailed FR for the probability distribution of SFI of forward and backward processes, granting the trajectories to represent an involution \cite{Crooks1999} dubbed detailed geometric FR. In Section \ref{sec:4}, we state our second main result, dubbed integral geometric FR, states that for an ensemble of trajectories between two equilibrium states, the accumulated entropy is, on average, upper bounded by the average variation of the thermal bath entropy. We emphasize that such results are not limited to near-equilibrium situations, but are in principle limited to stochastic trajectories between two equilibrium states. In Section \ref{sec: 5}, we derive an uncertainty relation for the information rate of the system in terms of the SFI that is analogous to a previously established uncertainty relation \cite{Nicholson2018}. To test the integral fluctuation relation, in Section \ref{sec:6} we apply our results for two simple models. Finally, in Section \ref{sec: 7} we present our conclusions. 


\medskip

\section{Stochastic Fisher information \label{sec:2}}

Within a stochastic setting, we will restrict, for simplicity, to the parameter case where $\theta(t) = t$, and it is possible to introduce the single-parameter time SFI~\cite{melo2024stochasticthermodynamicsfisherinformation}, 
\begin{equation}
    \iota_{f}(x,t) \equiv \left[\pdv{\log P(x,t)}{t}\right]^2,\label{eq:def_stochasticfi}
\end{equation}
which is directly related to stochastic entropy production. For the variation of the entropy of the system,
\begin{equation}
    s_{{\rm sys}}(x,t) \equiv -\ln P(x,t)
\end{equation}
leading to,
\begin{equation}
\Delta s_{\rm{sys}} = -\ln \frac{P(x_{f},t_{f})}{P(x_{0},0)}
\end{equation},
and therefore
\begin{equation}
\iota_{f}(x,t) = \left[\partial_{t}s_{\rm{sys}}(x,t)\right]^2
\end{equation}.
The definition of the SFI implies that for each instant $t$ and position $x$ it follows a distribution $P(\iota_{f}|x,t)$. In Ref.~\cite{melo2024stochasticthermodynamicsfisherinformation}, $P(\iota_{f}|x,t)$ was numerically calculated considering a particle under a driving nonequilibrium force. For an arbitrary $\theta$ parameter the mean SFI equals the original FI, $\langle \iota_{f}(x,t)\rangle_{P(x,t)} = \mathcal{I}_{F}(t)$.

In other words, since FI is a metric in the probability space, the SFI is a stochastic metric \cite{Moffat1997}. Informationally, it represents the full distribution of surprisal rate beyond just the average~\cite{shannon1948} by taking into account its fluctuations that occur when a given trajectory is performed; this means that $\mathcal{I} (t)$ is obtained from $\iota (x,t)$ by averaging over all trajectories. 
By taking into account the full distribution of $P(\iota_{f}|x,t)$, we find FRs of geometric nature.

\medskip

\section{Detailed FR for SFI \label{sec:3}}  

Consider $\iota_{f}(x,t)$ for a continuous distribution and assume that the time evolution is governed by a Markovian and stochastic process. Let us define $x(t)$ as the position of the particle at time $t$; $x(t)$ follows a distribution,
\begin{equation}
  P(x, t) = \int dx_{0}\rho(x_{0})\mathcal{P}[x(t)|x(0)] 
\end{equation}
and $[x(t)]_{t = 0}^{t = \tau}$ is a path that the system follows from $t = 0$ to $t = \tau > 0$. The entropy production a functional of this path. The total entropy produced is given by
\begin{equation}
    \Delta s_{{\rm tot}}[x(t)] = \Delta s_{{\rm sys}}[x(t)] + \beta q[x(t)],
\end{equation}
where $\Delta s_{{\rm sys}}[x(t)] = s_{{\rm sys}}(x_{f},t_{f}) -  s_{{\rm sys}}(x_{0},t_{0})$ is the variation of informational entropy of the system and $q[x(t)]$ is the heat that the system gains along the trajectory $[x(t)]$.  Assuming a system where both $\rho(x_0) = P(x(0),0)$ and $\rho(\Bar{x}_{\tau}) = P(x(\tau),\tau)$ are equilibrium distributions, the detailed Fluctuation Relation for the direct and inverse trajectories \cite{Crooks1999} states that,
\begin{equation}
    \frac{\rho(x_0)\mathcal{P}[x(t)|x(0)]}{\rho(\Bar{x}_{t})\mathcal{P}^{\dagger}[x(0)|x(t)]} = e^{s_{\rm{tot}}[x(t)]}.
    \label{eq:det_FT}
\end{equation}
The notation $\Bar{x}_{\tau}$ stands for the distribution of the initial state of the reverse trajectory. The fluctuation theorem stated in Eq.~(\ref{eq:det_FT}) indicates that the forward process and time-reversal process have probability distributions comparable to each other, with the proportionality term given by the produced entropy through the forward process. One of the key elements that guarantees the validity of Eq.~(\ref{eq:det_FT}) is that the final probability distribution of the forward process $\rho(x_{t})$ is equivalent to the initial distribution of the reverse process $\rho(\bar{x}_{t})$ and vice versa. We name this property involution \cite{peliti2021}. This means that entropy production is odd under time reversal $\Delta s_{{\rm tot}}^{\dagger}[\bar{x}(t)] = -\Delta s_{{\rm tot}}[x(t)]$ \cite{Crooks1999}. To achieve such property in the examples we consider systems with initial and final distributions of equilibrium.

Bearing in mind that $\iota_{f}(x,t)$ is a quantity defined at each space point, the SFI is recast as a function of the stochastic entropy $s_{\rm{sys}}(x)$, in the form
\begin{equation}
    \iota_{f}(x,t) = f(s_{\rm{sys}}(x,t)).
\end{equation}
%

\paragraph{Proposition I: Detailed Geometric Fluctuation Relation --}
Given the Crooks Fluctuation theorem holds, consider the probability $P_{F}(\iota_{f})$ of observing a particular set of $\{\iota_{f}(t)\}$ values in the forward trajectory. This distribution can be written in terms of a Dirac delta function averaged by the ensemble of forward paths,
\begin{eqnarray}
    P_{F}(\iota_{f}) = \langle\delta(\iota_{f}(x,t) - f(s_{\rm{sys}}(x,t)))\rangle_{F} \nonumber\\ \equiv \iiint dx_{0} dx_{t}\mathcal{D}x~ \rho(x_{0}) \mathcal{P}[x(t)|x(0)] \delta(\iota_{f}(x,t) - f(s_{\rm{sys}}(x,t))),
\end{eqnarray}
$f(s_{{\rm sys}}(x,t))$ is a function of $s_{{\rm sys}}(x,t)$ such that for any arbitrary $a(x,t)$, $f(a(x,t)) = \left[\partial_{t}~a(x,t)\right]^2$. The integrating factor $\mathcal{D}x$ represents the integration over the possible paths given it represents a stochastic dynamics.

Using the detailed FT in Eq. (\ref{eq:det_FT}) one gets
\begin{equation}
    \frac{P_F(\iota_{f}(x,t))}{P_{B}(\iota_{f}(x,t))} = e^{q[x(t)]/T + \ell[x(t)]}.
    \label{crooks_sfi}
\end{equation}
for, 
\begin{equation}
    \ell[x(t)] \equiv \int {\rm d}t~ \sqrt{\iota_{f}(x,t)},\label{eq:stochastic_length}
\end{equation}
is the stochastic length, \textit{i.e.} the accumulated system entropy for a single stochastic trajectory defined at each point \footnote{For details in the calculations of Eq.~(\ref{crooks_sfi}), we refer to \ref{AppendixB}.}. In that setting, and assuming involution holds, we obtain a detailed FR, now dubbed detailed geometric fluctuation relation.

\textbf{Proof.} See \ref{AppendixB}.

Proposition 1 states the following: Consider a system in contact with a heat reservoir. If the involution of the trajectory holds -- ie, the initial and final states are the same equilibrium states of both forward and backward trajectories ----, the detailed fluctuation theorem of entropy is valid. Concomitantly, we argue that there exists a geometric detailed fluctuation relation for the distribution probability of forward SFI and backward SFI that is proportional to the exponential of the entropy accumulated in the forward trajectory. Concerning Eq.~(\ref{eq:stochastic_length}), it tells us that the contribution of the system to the entropic cost of reverting the natural entropic pathway is given by a stochastic length, meaning the path has a fluctuating metric.

The concept of stochastic length arises from the calculations of the detailed fluctuation relation. However, it differs from analogous concepts such as the thermodynamic length $\mathcal{L}(t) \equiv \int {\rm d}t~\sqrt{\mathcal{I}(t)}$, as it is the integral of a stochastic metric. Moreover, the average stochastic length $\langle \ell\rangle(t) = \int {\rm d}t~\langle\sqrt{\iota_{f}(x,t)} \rangle$ is related to $\mathcal{L}(t)$ through the inequality $\mathcal{L}(t) \ge \langle \ell\rangle(t)$.

The physical interpretation of $\ell[x(t)]$ can be understood from the perspective of entropy production during a single trajectory. Given the SFI represents the squared rate of the system entropy production, $\iota_{f}(x,t) = \left[ \partial_{t}s_{{\rm sys}}(x,t)\right]^2$, $\ell[x(t)]$ can be rewritten as
\begin{equation}
    \ell[x(t)] = \int {\rm d}t~\sqrt{(\partial_{t}s_{{\rm sys}}(x,t))^2} = \int {\rm d}t~\abs{\partial_{t}s_{{\rm sys}}(x,t)}.
\end{equation}
The result above indicates that the stochastic length measures the accumulated entropy of the system over a stochastic trajectory and contrast with the variation of entropy of the system over a trajectory, that is indicated only the the difference $\Delta s_{{\rm sys}}[x(t)] = s_{{\rm sys}}(x_{f},t_{f}) - s_{{\rm sys}}(x_{0},t_{0})$. In geometrical terms, while $\Delta s_{{\rm sys}}[x(t)]$ defines the displacement in entropy space, $\ell[x(t)]$ relates to the distance performed in the same space.

\medskip
\section{Integral Geometric FR for the SFI \label{sec:4}}
For the average of the ensemble of stochastic lengths between two equilibrium points, we have:
\paragraph*{Proposition II: Integral Geometric Fluctuation Relation.}
By stating the integration of Eq.~(\ref{crooks_sfi}) we get
\begin{equation}
\int \mathcal{D}\iota_{f} P_{F}(\iota_{f}) e^{-(q[x(t)]/T + \ell[x(t)])} = \int \mathcal{D}\iota_{f} P_{B}(\hat{\iota}_{f}) = 1,
\end{equation}
meaning,
\begin{equation}
    \langle e^{-(\beta q[x(t)] + \ell[x(t)])} \rangle = 1,
    \label{eq:av_heat_and_length}
\end{equation}
which is the integral FR, dubbed integral geometric fluctuation relation. From Jensen inequality, its yields the bounding relation,
\begin{equation}
     \langle \ell[x(t)]\rangle_{P(x_t,t,x_0,0)} + \Delta S_{\rm{bath}}(t) \le 0,
    \label{eq:Integral_FT_inequality}
\end{equation}
for $\Delta S_{\rm{bath}}$ is the entropy that outflows from the system to the thermal reservoir.

Proposition II states that the average accumulated entropy of an ensemble of stochastic trajectories between two equilibrium points is bounded by the variation of entropy of the bath, where the accumulated entropy of the system is given in terms of a stochastic length. The average result makes sense, since on the thermodynamic limit, the possible entropic length is limited due to the bath entropy. As such, this is a re-statement of the second law of thermodynamics, in terms of entropic lengths. Regarding the fact that $\ell[x(t)] \ge 0$, it stems from the fact that $\iota_{f}(x,t)$ is strictly positive, very much like in kinematics speed is non-negative. This part of the inequality expression is tightly bound, as there should not be any negative (stochastic) lengths.

Even more, violations of the inequality in Eq.~(\ref{eq:Integral_FT_inequality}) about the right inequality reveal the probabilistic character of the second law, as stated for instance in Refs.~\cite{jarzynski1997, Crooks1999}, since thermodynamic lengths of the system that exceed the variation of entropy $\Delta S_{\rm{bath}}$ are only possible in terms of fluctuations, not in the average. 

There is also the relation between the average stochastic length to the thermodynamic length, defined in Eq.~(\ref{eq:thermodynamic_length}). In fact, one has that the average stochastic length is a convolution of the thermodynamic length, that is
\begin{equation}
  L(t) \equiv \langle \ell[x(t)]\rangle \le \mathcal{L}(t).  
\end{equation}
One can see that the above inequality leads to the relation $\langle\sqrt{\iota(x,t)}\rangle \le \sqrt{\langle \iota(x,t)\rangle}$, which means that the correspondent probability distribution that saturates the bound is a constant distribution. One of such is a equal \textit{a priori} microcanonical distribution. This suggests that $L(t)$ is equivalent to $\mathcal{L}(t)$ only in equilibrium.

\section{Stochastic Uncertainty Relations \label{sec: 5}} 
In analogy to references \cite{crooks2007, Ito2018}, we will introduce a stochastic measure of uncertainty over a path across the stochastic manifold $\Theta[x(t)]$. We want to show that we can also establish a bound for the uncertainty from the stochastic geodesic, on trajectory level. As known, the conventional FI from Eq.~(\ref{FI_average}) is an appropriated metric for the line element that represents a displacement between two distributions ${\rm d}s^2 = \mathcal{I}(t) {\rm d}t^{2}$. Likewise, we assume that the stochastic FI represents an appropriate metric for the stochastic line element that represents a displacement between two distributions ${\rm d}s^2(x,t) = \iota_{f}(x,t)~{\rm d}t^2$. As previously introduced, the length of a path in the stochastic manifold $\Theta[x(t)]$ is represented by $\ell[x(t)] = \int {\rm d}t \sqrt{\iota_{f}(x,t)}$. By making use of the Cauchy-Schwarz inequality, we get
\begin{equation}
    j[x(t)] \equiv \tau \int_{t_0}^{t_f}{\rm d}t~\iota_{f}(x,t) \ge \ell^2[x(t)], \label{eq:divergence_action}
\end{equation}
where $j[x(t)]$ is a stochastic action \cite{Heseltine_2016, Nicholson2016}, which is a trajectory-dependent quantity, it is a time integration of the SFI along a stochastic trajectory. Eq.~(\ref{eq:divergence_action}) represents an analogy to a previously established result, that shows that for an action $\mathcal{J} \equiv \tau\int{\rm d}t~ \mathcal{I}(t)$, the divergence $\mathcal{J} \ge \mathcal{L}^2$ holds \cite{crooks2007, Heseltine_2016, Nicholson2016}.

As argued in Refs.~\cite{Heseltine_2016, Nicholson2016, Nicholson2018}, the difference $\mathcal{J} - \mathcal{L}^2 \ge 0$ represents a temporal variance, meaning it corresponds to the cumulative fluctuations for irreversible processes,ie, the cumulative deviations from a geodesic path. In our work, we introduce a stochastic temporal variance to compute the cumulative fluctuations for a given trajectory also by defining the temporal average $\mathbb{E}[A(t)] = \frac{1}{\tau}\int{\rm d}t~A(t)$. We have then the time-averaged stochastic variance
\begin{equation}
    \sigma[x(t)] = \frac{j[x(t)] - \ell^{2}[x(t)]}{\tau^2}  = \mathbb{E}[\iota_{f}(x,t)] - \mathbb{E}[\sqrt{\iota_{f}(x,t)}]^2 \ge 0, 
\end{equation}
which denotes the cumulative deviations from the stochastic geodesic that connects the initial and final distributions.
On the other hand,
\begin{equation}
    \sigma^2 \equiv \langle j[x(t)]\rangle - \langle \ell[x(t)]\rangle^2 \ge 0
\end{equation}
is a temporal variance. In that fashion, we define here a stochastic temporal variance 
\begin{equation}
    \sigma^2 [x(t)] \equiv j[x(t)] - \ell^{2}[x(t)],
\end{equation}
such that, on average,
\begin{equation}
    \langle \sigma^2[x(t)]\rangle_{P(x_{t},t,x_0,0)} = \langle j[x(t)]\rangle - \langle \ell^2[x(t)]\rangle \ge \sigma^2.
\end{equation}

Following Ref. \cite{Nicholson2018}, one has,
\begin{equation}
   \sigma^2 \le \frac{1}{\tau}\int dt~(\mathcal{I}(t) + \langle\sqrt{\mathcal{I}(t)}\rangle_{\rm{t}}^2) \le \frac{2\mathcal{J}}{\tau^2} 
\end{equation},
where $\langle A(t)\rangle_{\rm{t}} \equiv \frac{1}{\tau}\int_{t_{0}}^{t_{f}} dt~A(t)$, and the time-averaged action $\mathcal{J} \equiv \langle \langle j[x(t)]\rangle\rangle_{t}$. From that inequality, one gets,
\begin{equation}
    \langle\mathcal{I}_{F}(t)\rangle_{\rm{t}}\sigma^{-2} \ge \frac{1}{2}.\label{eq:delCampo_ineq.}
\end{equation}

Moreover, taking the relation $\langle \sigma^{2}[x(t)]\rangle \ge \sigma^2 \ge 0$, another bound can be found when using the stochastic temporal variance, namely,
\begin{equation}
    \langle\sigma^2[x(t)]\rangle_{\rm{t}} \le \langle j[x(t)]\rangle_{\rm{t}} + \langle\ell^2[x(t)]\rangle_{\rm{t}} \le \frac{2\langle j[x(t)]\rangle_{\rm{t}}}{\tau^2},
\end{equation}
where $\langle \ell[x(t)]\rangle^2 \le \langle\ell^2[x(t)]\rangle_{\rm{t}}$. We can show an uncertainty relation for $\langle\sigma^{2}[x(t)]\rangle_{\rm{t}}$ and between it and $\sigma^2$. With the SFI, we can define $\sigma[x(t)]$ for a single trajectory, and on an trajectory ensemble average it holds Eq.~(\ref{eq:delCampo_ineq.}).

\medskip
\section{Examples\label{sec:6}}
\subsection{Example I: Particle subject to a saturating driving force} The Langevin Equation (LE) for a Brownian particle subject to a driving force that grows to a constant value is given by
\begin{equation}
 \gamma\Dot{x}(t) = -k x(t) + F_{0}(1 - e^{-t/\tau})x(t) + \eta(t),  
\end{equation}
where we consider white noise such that $\langle \eta(t)\eta(t')\rangle = 2\gamma\beta^{-1}\delta(t - t')$, with $\beta = T^{-1}$. This LE represents the dynamics of a particle in contact with a reservoir at temperature $T$ subjected to a restoring force $-k\, x(t)$ and under the influence of a driving force $F(t) = F_{0}(1 - e^{-t/\tau})$ \cite{melo2024stochasticthermodynamicsfisherinformation, morgado2010exact}. 

Via path integral methods~\cite{Wio_Path}, we calculate the conditional probability distribution for a Brownian particle to be in position $x(t)$ at time $t$ given it was localized in position $x(0)$ at time $t = 0$ \footnote{The complete expressions for the conditional probability distribution and the marginal probability distribution of $x_{t}$ are stated on \ref{AppendixA}}. Then, by considering an initial equilibrium state and considering that the system subject to the adopted driving force is driven to equilibrium in the infinite time limit \cite{melo2024stochasticthermodynamicsfisherinformation}, we calculate the average represented in Eq.~(\ref{eq:av_heat_and_length}). It results in
\begin{equation}
    \langle e^{-q[x(t)]/T - \ell[x(t)]}\rangle = 1,
\end{equation}
as expected.
\medskip

\subsection{Example II: Particle in a harmonic potential with a constant velocity driving force}
Considering a particle in a harmonic potential that is dislocated with constant velocity $v$,
\begin{equation}
    \gamma \dot{x}(t) = -k(x(t) - vt) + \eta(t).
    \label{Langevin_eq}
\end{equation}
This describes the dynamics of a Brownian particle subjected to a harmonic potential that translates with $v$ velocity, in contact with a Markovian heat reservoir at temperature $T$ \cite{vanZon2003}.

\begin{figure}
    \centering
    \includegraphics[width=1.0\linewidth]{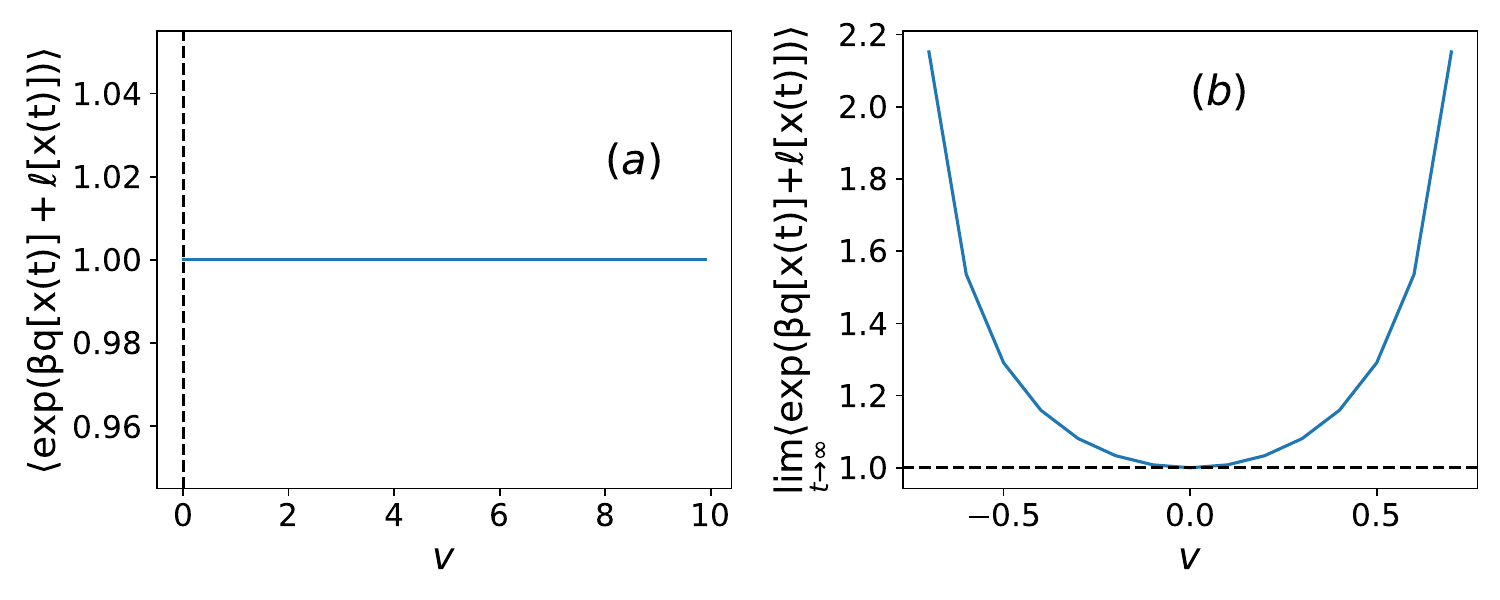}
    \caption{Integral Geometric Fluctuation Relations for the adopted examples with all system parameters but $v$ set to $1$. (a) Integral geometric FR for the Brownian particle subject to a saturating driving force. (b) Asymptotic time integral geometric FR for a Brownian particle in a harmonic potential that translates with constant velocity. In (b), the integral geometric FR is only valid in the quasi-static limit, meaning as $v$ grows the average of $\exp(\beta q[x(t)] + \ell[x(t)])$ deviates from $1$.}
    \label{fig:fig1}
\end{figure}

We obtain the transition probability $P[x_{t},t|x_{0},0]$ via path integral methods \cite{Wio_Path}. The joint probability of having positions $x_t$ at time $t$ and $x_0$ at $t=0$ is naturally  $P(x_{t},t;x_{0},0) = P[x_{t},t|x_{0},0]\rho_{0}(x_{0})$.
%
By calculating the average in Eq.~(\ref{eq:av_heat_and_length}), one gets, in the infinite time limit \footnote{The expression for the integral fluctuation relation without any limits taken is unnecessarily enormous due to its size. We emphasize that despite that, the infinite time limit is sufficient to sustain our arguments.}, the following result
\begin{equation}
    \lim_{t \rightarrow \infty}\langle e^{\beta q[x(t)] + \ell[x(t)]}\rangle = \sqrt{\frac{4 \gamma +1}{4 \gamma -8 \beta  \gamma  k v^2+1}},
\end{equation}
which, on the quasi-static limit $v \rightarrow 0$ tends to $1$ and deviates as $v$ grows. This result agrees with our prediction, as such a system only reaches the equilibrium in the quasi-static limit. Figure \ref{fig:fig1} displays the results of examples $I$ and $II$, for the general time and the asymptotic limit, respectively, with all parameters but $v$ set to $1$. For $(b)$ we observe that the integral geometric FR is only valid at the quasi-static limit, as expected.

\medskip 

\medskip
\section{Discussions and conclusions \label{sec: 7}}

In this paper, we have investigated the stochastic Fisher Information and derive two fluctuation relations within the framework of stochastic geometry, referred to as geometric fluctuation relations. This framework introduces a stochastic length based on individual non-equilibrium trajectories, offering insights into the accumulation of system entropy rather than just the difference between initial and final states.

The detailed geometric fluctuation relation states that the distribution of stochastic metrics (SFI) for reverse processes between two equilibrium states is proportional to that of forward processes, adjusted by an exponential factor of the total entropy along the forward trajectory.

The integral geometric fluctuation theorem asserts that, for an ensemble of possible non-equilibrium trajectories between two equilibrium states, the average stochastic length is bounded from below by the entropy variation that flows into the coupled thermal bath. In equilibrium, the average stochastic length is equal to the thermodynamic length of the path, defined in terms of the FI. There remains the question if there is a bound between the thermodynamic length and the bath entropy. 

Finally, we introduce an uncertainty relation for temporal stochastic uncertainty, analogous to those for temporal Fisher information. Using the SFI, we derive a single-trajectory uncertainty based on the stochastic action and length of the trajectory. This development paves the way to the formulation of thermodynamic uncertainty relations for ensembles and trajectory-dependent speed limits.

We address the differences between the geometric FRs and other known, apparently similar fluctuation relations, namely the Hatano-Sasa~\cite{hatano2001}, and Speck-Seifert~\cite{Seifert_2010}, integral fluctuation relations that are valid for systems in non-equilibrium steady states (NESS). In the NESS, the total heat produced can be written as excess and housekeeping heat~\cite{hatano2001}, and the response to any observable is proportional to the entropy production~\cite{Seifert_2010}. Contrarily to the aforementioned relations, the geometric FR we have introduced is valid for trajectories between two equilibrium states in probability space, providing previously inaccessible insight into the nonequilibrium dynamics between two equilibrium states. Another relevant point is the choice of the SFI as the adopted stochastic metric. In principle, the results uncovered here are valid for any consistent stochastic metric. For instance, geometric FRs can be derived for a stochastic version of the Jensen-Shannon divergence or any other metric in principle. The adoption of SFI is a mere practical question, as it seemed straightforward to us at first.

The relations derived in Eqs.~(\ref{crooks_sfi}) and~(\ref{eq:Integral_FT_inequality}) apply to any non-equilibrium trajectories between two equilibrium states. We believe that analysing the system's entropy content can provide new insights into the physical processes involved in system equilibration. Furthermore, the experimental reproducibility of these results appears feasible for systems in which time-dependent probability distributions can be accessed, such as optical tweezer experiments and colloidal particles~\cite{Cilibertoprx2017, martins2022thermodynamicmeasurementnonequilibriumstochastic, Luna2024, Kremer2024}. Among the possible extensions of this work are stochastic speed limits~\cite{Okuyama2018, Shiraishi2018, Ito2018, Gupta2020}, as the stochastic length opens an avenue of research for exploring trajectory-level speed limits and generalizations to another stochastic metrics.

\section*{Data Availability}
No new data were created or analysed in this study. Data sharing is not applicable to this article.

\section*{Acknowledgments}
PBM acknowledges Pedro V. Paraguass\'{u} for insightful discussions and Hammond Smith for useful insights. This work is supported by the Brazilian agencies Conselho Nacional de Desenvolvimento Científico e Tecnológico (CNPq), Coordena\c{c}\~{a}o de Aperfei\c{c}oamento de Pessoal de Ensino Superior (CAPES), Funda\c{c}\~{a}o Carlos Chagas de Apoio \`a Pesquisa do Estado do Rio de Janeiro (FAPERJ), and Funda\c{c}\~{a}o de Amparo \`a Pesquisa do Estado de S\~ao Paulo (FAPESP). PBM acknowledges CAPES' scholarship, finance code 001. SMDQ acknowledges CNPq grant No. 302348/2022-0.  SMDQ acknowledges FAPERJ grant No. APQ1-210.310/2024.DOSP acknowledges the Brazilian funding agencies CNPq (Grants No. 307028/2019-4, 402074/2023-8), FAPESP (Grant No. 2017/03727-0), and the Brazilian National Institute of Science and Technology of Quantum Information (INCT-IQ) Grant No. 465469/2014-0. WAMM acknowledges CNPq grant No. 308560/2022-1.

\appendix
\section{Probability distributions for the adopted examples\label{AppendixA}} We put here the implicit form of the probabilities calculated via path integral methods to make our calculations' reproducibility possible. 

For the first example, the conditional probability $\mathcal{P}[x(t)|x(0)]$ is given by
\begin{eqnarray}
   \mathcal{P}[x(t)|x(0)] = \frac{\sqrt{\beta  k \Theta_{+}(t)}}{2 \sqrt{\pi }}\exp \left(-a(x(0))x^2(t) - b(x(0))x(t) - c(x(0))\right),
\end{eqnarray}
with $\Theta_{\pm}(t) = \left(\coth \left(\frac{k t}{\gamma }\right)\pm 1\right)$, 

\begin{equation}a = \frac{1}{4} \beta  k e^{\frac{2 k t}{\gamma }} \Theta_{-}(t),
\end{equation}
\begin{equation}
     b = \frac{\beta  e^{\frac{k t}{\gamma }} \Theta_{-}(t)}{2 (\gamma -k \tau )} \left(\gamma  F_0-F_0 e^{\frac{k t}{\gamma }-\frac{t}{\tau }} \left(e^{t/\tau } (\gamma -k \tau )+k \tau \right)+k x_0 (k \tau -\gamma )\right),
\end{equation}
and 
\begin{equation}
c = \frac{\beta  e^{-\frac{2 t}{\tau }} \Theta_{-}(t)}{4 k (\gamma -k \tau )^2} \left(F_0 e^{\frac{kt}{\gamma }} \left(e^{t/\tau } (\gamma -k \tau )+k \tau \right)-e^{t/\tau } \left(\gamma  F_0+k x_0 (k \tau-\gamma )\right)\right)^2.
\end{equation}

Considering an initial equilibrium state at an average position $x_{0} = x(t = 0)$, one has the following marginal probability distribution function for the position $x_{t} \equiv x(t)$
\begin{equation}
    P(x_t,t) = \sqrt{\frac{2}{\Theta_{+}(t)}}e^{-a'x_t^2 - b'x_t - c},
\end{equation}
where $a' = \frac{1}{2}k\beta$, $b' = \frac{\beta  F_0 \left(\gamma +\gamma  \left(-e^{-\frac{k t}{\gamma }}\right)+k \tau  \left(e^{-\frac{t}{\tau}}-1\right)\right)}{\gamma -k \tau }$, and 
\begin{equation}
c' = \frac{\beta  F_0^2 e^{-2 t \left(\frac{k}{\gamma }+\frac{1}{\tau }\right)}}{2 k (\gamma -k \tau )^2} \left(k \tau  e^{\frac{k t}{\gamma}}+(\gamma -k \tau ) e^{t \left(\frac{k}{\gamma }+\frac{1}{\tau }\right)}+\gamma  \left(-e^{t/\tau}\right)\right)^2.
\end{equation}

For the second example, the conditional probability distribution is given by
\begin{equation}
    \mathcal{P}[x(t)|x(0)] = \frac{\sqrt{k\beta \Theta_{+}(t)}}{4\pi}e^{-a''x^2(t) - b''x(t) - c''},
\end{equation}
where $a'' = \frac{1}{4}e^{\frac{2kt}{\gamma}}k\beta\Theta_{-}(t)$, $b'' = -\frac{1}{2} \beta  e^{\frac{k t}{\gamma }} \Theta_{-}(t) \left(v
   \left(\gamma +e^{\frac{k t}{\gamma }} (k t-\gamma )\right)+k x_0\right)$, and $c'' = -\frac{\beta  \Theta_{-}(t) \left(v \left(\gamma +e^{\frac{k t}{\gamma
   }} (k t-\gamma )\right)+k x_0\right)^2}{4 k}$.
   
Finally, the marginal PDF for $x_t = x(t)$ is given by
\begin{equation}
    P(x_t,t) = \sqrt{\frac{k\beta}{2\pi}}e^{-a'''x_t^2 + b'''x_t + c'''},
\end{equation}
with $a''' = \frac{1}{2}k\beta$, $b''' = \beta  v \left(\gamma -\gamma  e^{-\frac{k t}{\gamma }}-k t\right)$, and $c''' = \frac{\beta  v^2 e^{-\frac{2 k t}{\gamma }} \left(\gamma +e^{\frac{k t}{\gamma }} (k t-\gamma )\right)^2}{2 k}$.
\section{Details on the derivation of the geometric FR\label{AppendixB}}
Assuming a system where both $\rho(x_0) = P(x,0)$ and $\rho_{B}(\Bar{x}_{\tau}) = P(x,\tau)$ are equilibrium distributions; the detailed (Crooks) Fluctuation Theorem for the direct and inverse trajectories states
\begin{equation}
    \frac{\rho(x_0)\mathcal{P}[x(t)|x(0)]}{\rho(\Bar{x}_{t})\mathcal{P}^{\dagger}[x(0)|x(t)]} = e^{s_{\rm{tot}}[x(t)]}.
\end{equation}
By rewriting $\iota_{F}(x,t)$ as a trajectory dependent quantity,
\begin{equation}
    \iota_{f}(x,t) = (\partial_{t}\ln P(x,t))^2,
\end{equation}
the SFI is recast as a function of the system stochastic entropy $s_{\rm{sys}}(x)$, in the form
\begin{equation}
    \iota_{f}(x,t) = f(s_{\rm{sys}}(x,t)),~f(\cdot)\equiv(\partial_{t}\cdot)^2
\end{equation}

Consider now the probability $P_{F}(\iota_{\theta})$ of observing a particular value of $\iota_{\theta}$ in the forward trajectory; that distribution can be written in terms of a $\delta$-Dirac averaged by the ensemble of forward paths,
\begin{eqnarray}
    P_{F}(\iota_{f}) = \langle\delta(\iota_{f} - f(s_{\rm{sys}}(x,t)))\rangle_{F} \nonumber\\= \iiint dx_{0} dx_{t}\mathcal{D}x~ \rho(x_{0}) \mathcal{P}[x(t)|x(0)] \delta(\iota_{f} - f(s_{\rm{sys}}(x,t))).\label{eq:det_FT2}
\end{eqnarray}

Using the detailed FT in Eq. (\ref{eq:det_FT}) one gets
\begin{eqnarray}
     P_{F}(\iota_{f}) = \langle\delta(\iota_{f} - f(s_{\rm{sys}}(x,t)))\rangle_{F}\nonumber\\ = \iiint dx_{0}~dx_{t}~\mathcal{D}x  \rho(\Bar{x}_{t})e^{s_{\rm{tot}}} \delta(\iota_{f} - f(s_{\rm{sys}}(x,t)))\mathcal{P}^{\dagger}[x(0)|x(t)].
\end{eqnarray}
By changing variables of the $\delta$ function between $\iota_{f}(x,t)$ and $s_{{\rm sys}}(x,t)$, we get
\begin{eqnarray}
     P_{F}(\iota_{f}) = \iiint dx_{0}~dx_{t}~\mathcal{D}x \abs{\dv{s_{{\rm sys}}(x,t)}{\iota_{f}(x,t)}}^{-1} \rho(\Bar{x}_{t})e^{s_{\rm{tot}}} \mathcal{P}^{\dagger}[x(0)|x(t)]\nonumber\\\times \delta(s_{\rm{sys}} - \ell[x(t)]),
\end{eqnarray}
and given $\Delta s_{{\rm sys}} = \Delta s_{{\rm tot}} - q[x(t)]/T$, we substitute the $\delta$ function terms into $\delta(s_{{\rm tot}}-q[x(t)]/T - \ell[x(t)])$, to obtain the exponential term in terms of heat and the $\ell[x(t)] \equiv f^{-1}(\iota_{f})$ quantity, meaning the stochastic length, and make the change of variables of the delta function from $s_{{\rm sys}}$ to $\iota_{f}$ again
\begin{eqnarray}
    P_{F}(\iota_{f}) = e^{q[x(t)] + \ell[x(t)]}\iiint dx_{0}~dx_{t}~\mathcal{D}x~ \rho(\Bar{x}_{t}) \mathcal{P}^{\dagger}[x(0)|x(t)] \nonumber\\\times\delta(\iota_{f} - f(s_{{\rm sys}}(x,t))).
\end{eqnarray}
The emergent $\ell[x(t)] \equiv f^{-1}(\iota_{F}(x,t))$ quantity, called stochastic length, defined by
\begin{equation}
    \ell[x(t)] \equiv \int dt~ \sqrt{\iota_{f}(x,t)},\label{eq:stochastic_length2}
\end{equation}
represents the accumulated system entropy for a single stochastic trajectory defined at each point. $\iota_{f}(x,t)$ is a random number that is given at each point making $\ell[x(t)]$ a trajectory-dependent variable.

\section*{References}
\bibliographystyle{name.bst}
\bibliography{refs.bib}

\providecommand{\newblock}{}
\begin{thebibliography}{10}
\expandafter\ifx\csname url\endcsname\relax
  \def\url#1{{\tt #1}}\fi
\expandafter\ifx\csname urlprefix\endcsname\relax\def\urlprefix{URL }\fi
\providecommand{\eprint}[2][]{\url{#2}}

\bibitem{Evans1993}
Evans D~J, Cohen E~G~D and Morriss G~P 1993 {\em Phys. Rev. Lett.\/} {\bf 71}(15) 2401--2404 \urlprefix\url{https://link.aps.org/doi/10.1103/PhysRevLett.71.2401}

\bibitem{Gallavoti1995}
Gallavotti G and Cohen E~G~D 1995 {\em Phys. Rev. Lett.\/} {\bf 74}(14) 2694--2697 \urlprefix\url{https://link.aps.org/doi/10.1103/PhysRevLett.74.2694}

\bibitem{jarzynski1997}
Jarzynski C 1997 {\em Phys. Rev. Lett.\/} {\bf 78} 2690

\bibitem{Crooks1999}
Crooks G~E 1999 {\em Phys. Rev. E\/} {\bf 60}(3) 2721--2726 \urlprefix\url{https://link.aps.org/doi/10.1103/PhysRevE.60.2721}

\bibitem{Weinhold1975}
Weinhold F 1975 {\em The J. of Chem. Phys.\/} {\bf 63} 2479--2483 ISSN 0021-9606 \urlprefix\url{https://doi.org/10.1063/1.431689}

\bibitem{Ruppeiner1979}
Ruppeiner G 1979 {\em Phys. Rev. A\/} {\bf 20}(4) 1608--1613 \urlprefix\url{https://link.aps.org/doi/10.1103/PhysRevA.20.1608}

\bibitem{Salamon1983}
Salamon P and Berry R~S 1983 {\em Phys. Rev. Lett.\/} {\bf 51}(13) 1127--1130 \urlprefix\url{https://link.aps.org/doi/10.1103/PhysRevLett.51.1127}

\bibitem{Ruppeiner1995}
Ruppeiner G 1995 {\em Rev. Mod. Phys.\/} {\bf 67}(3) 605--659 \urlprefix\url{https://link.aps.org/doi/10.1103/RevModPhys.67.605}

\bibitem{Brody1995}
Brody D and Rivier N 1995 {\em Phys. Rev. E\/} {\bf 51}(2) 1006--1011 \urlprefix\url{https://link.aps.org/doi/10.1103/PhysRevE.51.1006}

\bibitem{crooks2007}
Crooks G~E 2007 {\em Phys. Rev. Lett.\/} {\bf 99} 100602

\bibitem{frank2009}
Frank S~A 2009 {\em J. Evol. Biology\/} {\bf 22} 231

\bibitem{mayer2006}
Mayer A~L, Pawlowski C~W and Cabezas H 2006 {\em Ecological Modelling\/} {\bf 195} 72

\bibitem{sahalia2008}
A\"it-Sahalia Y and Jacod J 2008 {\em Econometrica\/} {\bf 76} 727

\bibitem{Escher2011}
Escher B~M, de~Matos~Filho R~L and Davidovich L 2011 {\em Nature Physics\/} {\bf 7} 406--411 ISSN 1745-2481 \urlprefix\url{https://doi.org/10.1038/nphys1958}

\bibitem{Toth2012}
T\'oth G 2012 {\em Phys. Rev. A\/} {\bf 85}(2) 022322 \urlprefix\url{https://link.aps.org/doi/10.1103/PhysRevA.85.022322}

\bibitem{hyllus2012}
Hyllus P, Laskowski W, Krischek R, Schwemmer C, Wieczorek W, Weinfurter H, Pezzé L and Smerzi A 2012 {\em Phys. Rev. A\/} {\bf 85} 022321

\bibitem{Liu_2020}
Liu J, Yuan H, Lu X~M and Wang X 2019 {\em Journal of Physics A: Mathematical and Theoretical\/} {\bf 53} 023001 \urlprefix\url{https://dx.doi.org/10.1088/1751-8121/ab5d4d}

\bibitem{min2022}
Yu M, Liu Y, Yang P, Gong M, Cao Q, Zhang S, Liu H, Heyl M, Ozawa T, Goldman N and Cai J 2022 {\em npj Quant. Inf.\/} {\bf 8} 56

\bibitem{marvian2022}
Marvian I 2022 {\em Phys. Rev. Lett.\/} {\bf 129} 190502

\bibitem{Cramer_1946}
Cramér H 1999 {\em Mathematical Methods of Statistics (PMS-9)\/} (Princeton University Press) ISBN 9780691005478 \urlprefix\url{http://www.jstor.org/stable/j.ctt1bpm9r4}

\bibitem{Rao1992}
Rao C~R 1992 {\em Information and the Accuracy Attainable in the Estimation of Statistical Parameters\/} (New York, NY: Springer New York) pp 235--247 ISBN 978-1-4612-0919-5 \urlprefix\url{https://doi.org/10.1007/978-1-4612-0919-5_16}

\bibitem{Nicholson2018}
Nicholson S~B, del Campo A and Green J~R 2018 {\em Phys. Rev. E\/} {\bf 98}(3) 032106 \urlprefix\url{https://link.aps.org/doi/10.1103/PhysRevE.98.032106}

\bibitem{Ito2018}
Ito S 2018 {\em Phys. Rev. Lett.\/} {\bf 121}(3) 030605 \urlprefix\url{https://link.aps.org/doi/10.1103/PhysRevLett.121.030605}

\bibitem{TanvanVU2019_TUR}
Hasegawa Y and Van~Vu T 2019 {\em Phys. Rev. E\/} {\bf 99}(6) 062126 \urlprefix\url{https://link.aps.org/doi/10.1103/PhysRevE.99.062126}

\bibitem{ItoDechant2020}
Ito S and Dechant A 2020 {\em Phys. Rev. X\/} {\bf 10}(2) 021056 \urlprefix\url{https://link.aps.org/doi/10.1103/PhysRevX.10.021056}

\bibitem{nicholson2020}
Nicholson S, Garc\'ia-Pintos L, del Campo A and Greem J 2020 {\em Nature Phys.\/} {\bf 16} 1211

\bibitem{Falasco2020}
Falasco G and Esposito M 2020 {\em Phys. Rev. Lett.\/} {\bf 125}(12) 120604 \urlprefix\url{https://link.aps.org/doi/10.1103/PhysRevLett.125.120604}

\bibitem{Busch2002}
Busch P 2002 {\em The Time-Energy Uncertainty Relation\/} (Berlin, Heidelberg: Springer Berlin Heidelberg) pp 69--98 ISBN 978-3-540-45846-3 \urlprefix\url{https://doi.org/10.1007/3-540-45846-8_3}

\bibitem{Okuyama2018}
Okuyama M and Ohzeki M 2018 {\em Phys. Rev. Lett.\/} {\bf 120}(7) 070402 \urlprefix\url{https://link.aps.org/doi/10.1103/PhysRevLett.120.070402}

\bibitem{Shiraishi2018}
Shiraishi N, Funo K and Saito K 2018 {\em Phys. Rev. Lett.\/} {\bf 121}(7) 070601 \urlprefix\url{https://link.aps.org/doi/10.1103/PhysRevLett.121.070601}

\bibitem{Gupta2020}
Gupta D and Busiello D~M 2020 {\em Phys. Rev. E\/} {\bf 102}(6) 062121 \urlprefix\url{https://link.aps.org/doi/10.1103/PhysRevE.102.062121}

\bibitem{TanVanVu2020}
Vo V~T, Van~Vu T and Hasegawa Y 2020 {\em Phys. Rev. E\/} {\bf 102}(6) 062132 \urlprefix\url{https://link.aps.org/doi/10.1103/PhysRevE.102.062132}

\bibitem{Salazar2022}
Salazar D~S~P 2022 {\em Phys. Rev. E\/} {\bf 106}(3) L032101 \urlprefix\url{https://link.aps.org/doi/10.1103/PhysRevE.106.L032101}

\bibitem{TanVanVu2023}
Van~Vu T and Saito K 2023 {\em Phys. Rev. X\/} {\bf 13}(1) 011013 \urlprefix\url{https://link.aps.org/doi/10.1103/PhysRevX.13.011013}

\bibitem{melo2025_speedlimits}
Melo P~B, Iemini F, Duarte~Queir\'os S~M and Morgado W~A~M 2025 {\em In preparation.\/}

\bibitem{melo2024stochasticthermodynamicsfisherinformation}
Melo P~B, Duarte~Queir\'os S~M and Morgado W~A~M 2025 {\em Phys. Rev. E\/} {\bf 111}(1) 014101 \urlprefix\url{https://link.aps.org/doi/10.1103/PhysRevE.111.014101}

\bibitem{Moffat1997}
Moffat J~W 1997 {\em Phys. Rev. D\/} {\bf 56}(10) 6264--6277 \urlprefix\url{https://link.aps.org/doi/10.1103/PhysRevD.56.6264}

\bibitem{seifert2005}
Seifert U 2005 {\em Phys. Rev. Lett.\/} {\bf 95} 040602

\bibitem{shannon1948}
Shannon C~E 1948 {\em Bell Syst. Tech. J.\/} {\bf 27} 379--423

\bibitem{peliti2021}
Peliti L and Pigolotti S 2021 {\em Stochastic Thermodynamics: An Introduction\/} (Princeton -- NJ: Princeton University Press)

\bibitem{Heseltine_2016}
Heseltine J and jin Kim E 2016 {\em J. Phys. A\/} {\bf 49} 175002 \urlprefix\url{https://dx.doi.org/10.1088/1751-8113/49/17/175002}

\bibitem{Nicholson2016}
Nicholson S and Kim E~j 2016 {\em Entropy\/} {\bf 18} ISSN 1099-4300 \urlprefix\url{https://www.mdpi.com/1099-4300/18/7/258}

\bibitem{morgado2010exact}
Morgado W and Soares-Pinto D 2010 {\em Phys. Rev. E\/} {\bf 82} 021112

\bibitem{Wio_Path}
Wio H~S 2013 {\em Path Integrals for Stochastic Processes\/} (Singapore: World Scientific)

\bibitem{vanZon2003}
van Zon R and Cohen E~G~D 2003 {\em Phys. Rev. Lett.\/} {\bf 91}(11) 110601 \urlprefix\url{https://link.aps.org/doi/10.1103/PhysRevLett.91.110601}

\bibitem{hatano2001}
Hatano T and Sasa S~i 2001 {\em Phys. Rev. Lett.\/} {\bf 86} 3463

\bibitem{Seifert_2010}
Seifert U and Speck T 2010 {\em Europhys. Lett.\/} {\bf 89} 10007 \urlprefix\url{https://dx.doi.org/10.1209/0295-5075/89/10007}

\bibitem{Cilibertoprx2017}
Ciliberto S 2017 {\em Phys. Rev. X\/} {\bf 7}(2) 021051 \urlprefix\url{https://link.aps.org/doi/10.1103/PhysRevX.7.021051}

\bibitem{martins2022thermodynamicmeasurementnonequilibriumstochastic}
Martins T~T, Kamizaki L~P and Muniz S~R 2022 Thermodynamic measurement of non-equilibrium stochastic processes in optical tweezers (\textit{Preprint} \eprint{2209.05606}) \urlprefix\url{https://arxiv.org/abs/2209.05606}

\bibitem{Luna2024}
da~Fonseca A~L, Diniz K~a, Monteiro P~B, Pires L~B, Moura G~T, Borges M, Dutra R~S, Ether D~S, Viana N~B and Neto P~A~M 2024 {\em Phys. Rev. Res.\/} {\bf 6}(2) 023226 \urlprefix\url{https://link.aps.org/doi/10.1103/PhysRevResearch.6.023226}

\bibitem{Kremer2024}
Kremer O, Califrer I, Tandeitnik D, von~der Weid J~P, Tempor\~ao G and Guerreiro T 2024 {\em Phys. Rev. Appl.\/} {\bf 22}(2) 024010 \urlprefix\url{https://link.aps.org/doi/10.1103/PhysRevApplied.22.024010}

\end{thebibliography}

\end{document}